\documentclass[journal]{IEEEtran}
\ifCLASSINFOpdf
\else
\fi
\usepackage{amsmath}
\usepackage{amssymb}
\usepackage{graphicx}

\usepackage{textgreek}

\usepackage{gensymb}

\graphicspath{ {./figures_processed/} }

\begin{document}

%
\title{Grating design methodology for\\tailored free-space beam-forming}
%
%
%

\author{Gillenhaal~J.~Beck, Jonathan~P.~Home, and~Karan~K.~Mehta%
\thanks{G.J. Beck and J.P. Home are with the Institute of Quantum Electronics, ETH Zurich, Zurich, Switzerland. K.K. Mehta is with the school of Electrical and Computer Engineering, Cornell University, Ithaca, NY, USA. email: beckgi@phys.ethz.ch.}}%
\maketitle


\begin{abstract}
We present a design methodology for free-space beam-forming with general profiles from grating couplers which avoids the need for numerical optimization, motivated by applications in ion trap physics.
We demonstrate its capabilities through a variety of gratings using different wavelengths and waveguide materials, designed for new ion traps with all optics fully integrated, including UV and visible wavelengths.
We demonstrate designs for diffraction-limited focusing without restriction on waveguide taper geometry, emission angle, or focus height, as well as focused higher order Hermite-Gaussian and Laguerre-Gaussian beams.
Additional investigations examine the influence of grating length and taper angle on beam-forming, indicating the importance of focal shift in apertured beams.
The design methodology presented allows for efficient design of beam-forming gratings with the accuracy as well as the flexibility of beam profile and operating wavelength demanded by application in atomic systems.

\end{abstract}


%
\IEEEpeerreviewmaketitle

\section{Introduction}
\IEEEPARstart{W}{aveguide}-to-free-space outcoupling has enabled new developments in optical-phased arrays, beam-steering, LiDAR, and quantum information processing \cite{Poulton_Yaacobi_Cole_Byrd_Raval_Vermeulen_Watts_2017, Rhee_You_Yoon_Han_Kim_Lee_Kim_Park_2020, Poulton_Byrd_Russo_Timurdogan_Khandaker_Vermeulen_Watts_2019, Mehta_Bruzewicz_McConnell_Ram_Sage_Chiaverini_2016}, with high outcoupling efficiencies and straightforward fabrication making diffractive grating outcouplers ideal for applications requiring small device footprints and precise beam delivery.
In trapped-ion systems, integrated beam delivery in surface traps provides a number of benefits over free-space addressing including robustness to external vibrations, tight focusing, and the potential for scalability \cite{Mehta_Bruzewicz_McConnell_Ram_Sage_Chiaverini_2016}. 
Systems for various ion species have been demonstrated \cite{Niffenegger_Stuart_SoraceAgaskar_Kharas_Bramhavar_Bruzewicz_Loh_Maxson_McConnell_Reens_etal_2020, Ivory_Setzer_Karl_McGuinness_DeRose_Blain_Stick_Gehl_Parazzoli_2021} as well as high-fidelity entanglement \cite{Mehta_Zhang_Malinowski_Nguyen_Stadler_Home_2020}, indicating promise for scalable trapped-ion systems with applications from quantum sensing and metrology to large-scale quantum computing \cite{Moody_Sorger_Blumenthal_Juodawlkis_Loh_Sorace_Agaskar_Jones_Balram_Matthews_Laing_et_al_2022}.

Grating coupler design methodologies are most commonly motivated by efficient coupling to fibers, often for near-infrared wavelengths \cite{Waldhauesl_1997, Nadovich_Jemison_Kosciolek_Crouse_2017, VanLaere_Claes_Schrauwen_Scheerlinck_Bogaerts_Taillaert_OFaolain_VanThourhout_Baets_2007, Roelkens_VanThourhout_Baets_2006, Zou_Yu_Ye_Liu_Deng_Zhang_2015, Zhang_Li_Tu_Song_Zhou_Luo_Huang_Yu_Lo_2014}.
More recently, devices targeting free-space emission have been presented for various applications in atomic systems \cite{Kim_Westly_Roxworthy_Li_Yulaev_Srinivasan_Aksyuk_2018, Massai_Schatteburg_Home_Mehta_2022}, but current methodologies involve approximations that pose challenges when faced with the stringent demands of these systems, including varied beam waist requirements, operation at multiple wavelengths spanning the UV to IR, and the delivery of nontrivial spatial field profiles. 

A general approach for grating chirp and apodization was demonstrated in \cite{Mehta_Ram_2017}, but grating line curvatures for transverse focusing were restricted to back-emitting geometries and determined according to approximations limiting focusing accuracy. 
Prior approaches have used analytical expressions for Gaussian beams \cite{ Mehta_Ram_2017, Oton_2016}, but lack generality and explicit consideration of top oxide cladding layers, where refractive distortion can substantially alter the gratings required for precise free-space beam forming.
Also commonplace are approximations for the effective index in the grating, assuming it either remains constant despite longitudinal apodization \cite{Mehta_Ram_2017, Khan_Combrie_DeRossi_2020} or follows a weighted average of the core/cladding indices \cite{Marchetti_Lacava_Khokhar_Chen_Cristiani_Richardson_Reed_Petropoulos_Minzioni_2017}.

Here, we present a unified, accurate, and flexible outcoupler design process which overcomes these constraints.
After bolstering the longitudinal chirp and apodization of \cite{Mehta_Ram_2017} with explicit integration of fabrication limits, we expand on the holographic grating line extraction of \cite{Oton_2016} with (1) a more general analytical treatment of the free-space beam and consideration of oxide cladding, (2) explicit accounting for the grating's varying effective index and the corresponding impact on the propagating taper field enabling accurate application in high-index-contrast platforms, and (3) the inclusion of a quartic term when expressing the grating lines in polynomial form.
Furthermore, we note the influence of Gaussian focal shift on grating-outcoupled beams (previously discussed only in context of lenses \cite{Li_1992}) and extend the theory to accurately inform design.
The resulting process rapidly produces designs for diffraction-limited tailored beam profiles at flexibly chosen emission angles and focus locations.

As illustrative examples, we present the results from fully vectorial 3D finite-difference-time-domain (FDTD) simulations of selected outcouplers from newly designed surface ion traps with fully integrated optics (currently in fabrication). 
These include micron-scale focusing of backward- and forward-emission with different waveguide materials, wavelengths, and focal heights, as well as tightly focused Hermite-Gaussian and Laguerre-Gaussian beams, realizing focal positions with $\sim$1 \textmugreek m accuracy.

\begin{figure}[!t]
\centering
\includegraphics[width=3.3in]{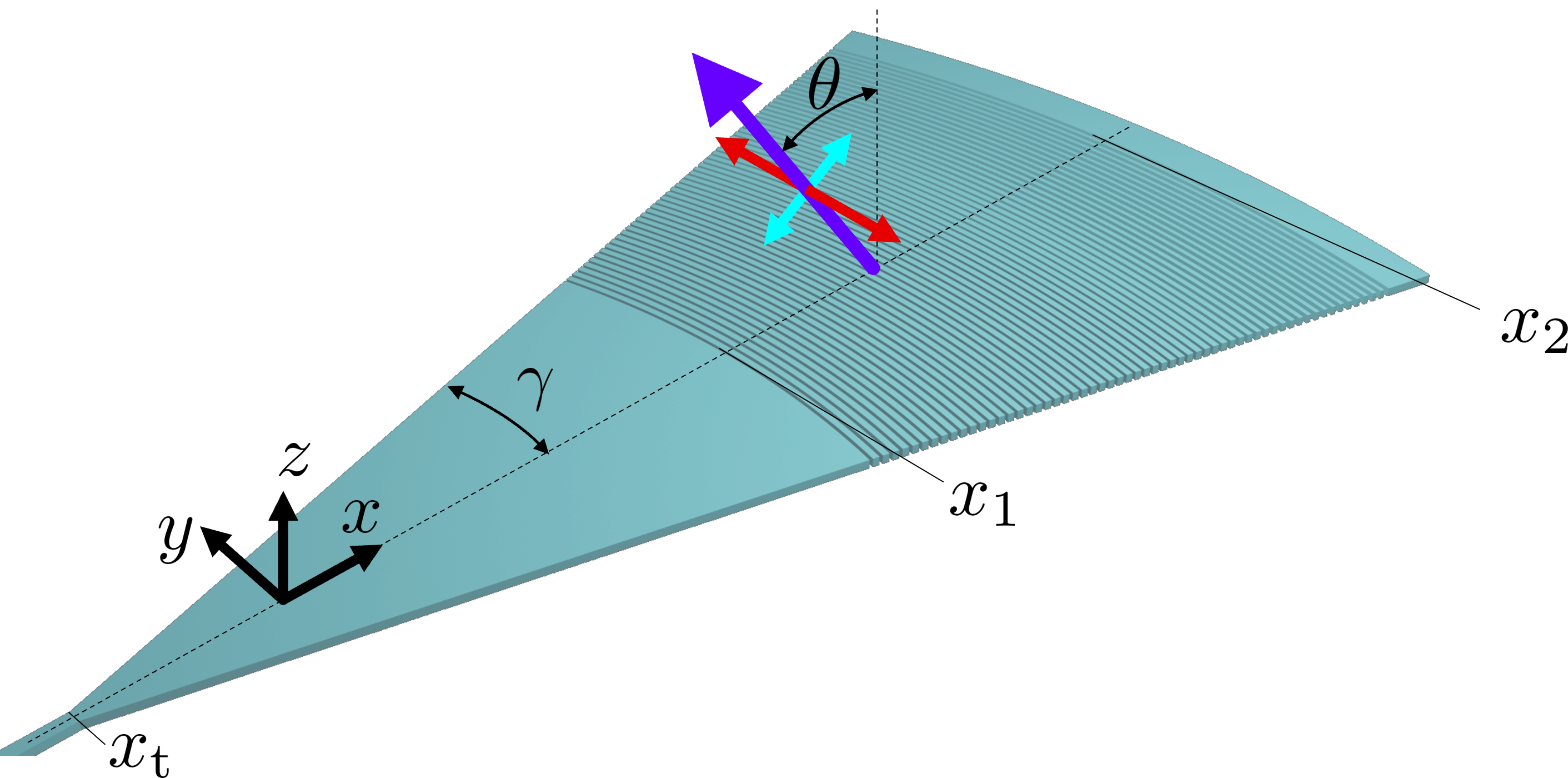}
\caption{Basic outcoupler and coordinate frame. 
Light enters from the waveguide along the $x$-axis and upon diffraction at the grating is emitted at angle $\theta$ (its positive direction being the ``back-emission'' shown).
Red and cyan arrows indicate the transverse and longitudinal beam axes (aligned with the $y$-axis and in the $xz$-plane, respectively).
We let the $x=0$ origin be directly below the target (ion) position; $x_\text{t}$ indicates the position of the taper start; $x_1$ and $x_2$ the grating start/end positions; and $\gamma$ the taper angle.
}
\label{fig:basic3d}
\end{figure}

\section{Grating design process}

Following the approach of \cite{Mehta_Ram_2017}, we start with ``longitudinal" design by considering the 2D cross-section along the grating centerline ($xz$-plane in Fig.~\ref{fig:basic3d}).
The emitted light is tailored through the angle of emission $\theta$ and the grating strength $\alpha$ (defined such that the guided amplitude decays as $e^{-\alpha x}$ for the position $x$ along the grating).
In a structure with constant etch depth, $\theta$ and $\alpha$ are determined by the period $\Lambda$ and duty cycle $DC$ (ratio of etch length to period).

After the 2D grating structure is set, we extend the gratings to 3D by extruding the perturbations along specified grating line curvatures.
The curvatures ensure phase matching between the waveguide and emitted fields, and can be determined from the holographic interference between the two.
Extraction of the in-grating effective index $n_\text{eff}(\Lambda, DC)$ from the longitudinal behavior allows accurate description of the propagating field's phase in the grating.
A specific outcoupled beam is thus produced by varying $\theta$ and $\alpha$ along the grating length to tailor the longitudinal focusing and beam intensity profile, and then considering the corresponding $n_\text{eff}$ values when utilizing the holographic interference to define the transverse structure for focusing and any additional phase profile (e.g. for higher-order modes).

The design process is broken down into distinct steps:
\begin{enumerate}
    \item determine the material stack and grating structures (informed by fabrication approach, e.g. etch depth, angles, etc.).
    \item determine $\alpha$, $\theta$, and $n_\text{eff}$ as a function of $\Lambda$ and $DC$.
    \item extract the design-relevant details of the desired emitted beam; in particular, the local angle of emission and intensity at the chip-vacuum interface and the phase in the waveguide plane.
     \item for longitudinal focusing and beam intensity profile: using the 2D simulation data, determine the local grating structural parameters ($\Lambda(x)$ and $DC(x)$) which correspond to the desired $\alpha(x)$ and $\theta(x)$ of the emitted beam. 
    \item for transverse focusing and phase profile: extract the form of the curved grating lines from the interference pattern between the fields of the desired emitted beam and the mode propagating from the waveguide.
\end{enumerate}

We have applied this methodology to wavelengths from the UV to IR and a wide range of beam emission angles, focuses, and ellipticities.
For demonstration, we follow the design process for a single device at 732 nm.

\subsection{Material stack and fabrication}

\begin{figure}[!t]
\centering
\includegraphics[width=3.3in]{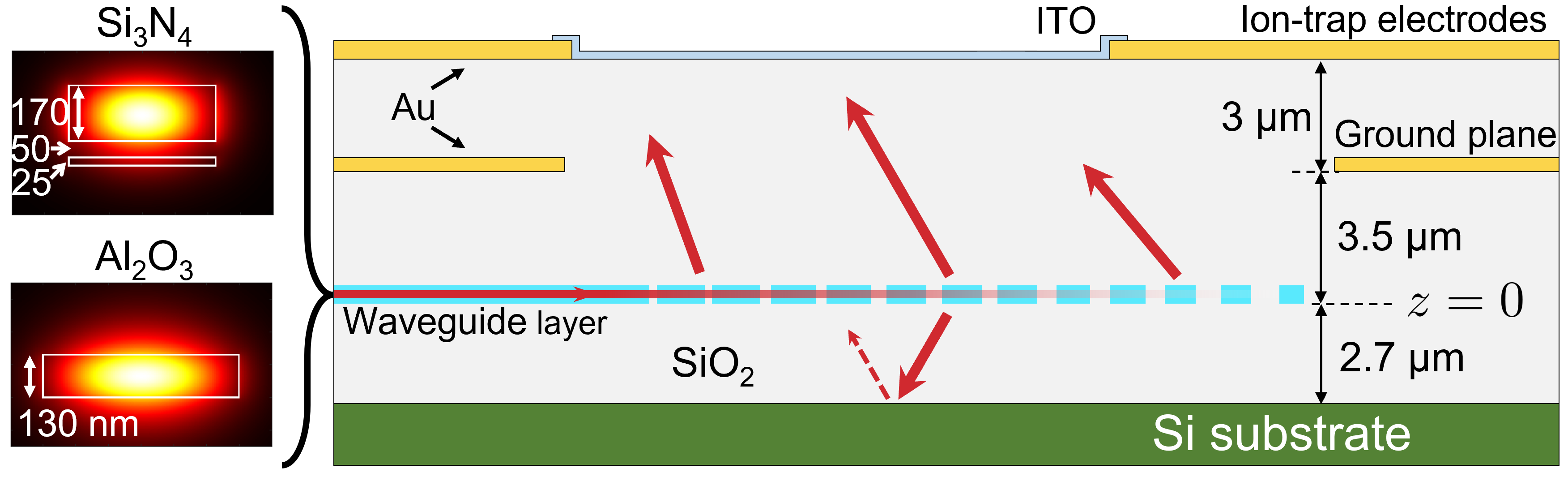}
\caption{
Device cross section illustrating backward-emitting diffraction (not to scale).
Waveguide layers consist of either asymmetric double-stripe Si$_3$N$_4$ or single-layer Al$_2$O$_3$.
Fundamental quasi-TE mode intensity profiles of waveguide cross sections are shown to the left; waveguide widths were typically 400-600 nm for single-mode operation (larger for higher-order waveguide modes).
The ion-trap chips include multiple electrode layers with cutouts for beam emission, as well as a thin (20 nm) ITO layer at the top oxide-vacuum interface, but these features do not impact the outcoupler design and are not included in the simulations presented.
}
\label{fig:material_stack}
\end{figure}

Designs discussed here were implemented within the layer stack shown in Fig.~\ref{fig:material_stack}, incorporating Al$_2$O$_3$ alongside Si$_3$N$_4$ to enable integration down to $\lambda=370$ nm \cite{West_Loh_Kharas_Sorace-Agaskar_Mehta_Sage_Chiaverini_Ram_2019}.
The waveguide layers sit on 2.7 \textmugreek m of thermal oxide (SiO$_2$) atop a silicon substrate.
A single 130 nm thick Al$_2$O$_3$ layer is used, whereas the Si$_3$N$_4$ utilizes an asymmetric double stripe structure with bottom/top layers of 25/170 nm separated by 50 nm of oxide (as in the previous generation \cite{Mehta_Zhang_Miller_Home_2019}).
Here we consider only quasi-TE modes, but the design process easily extends to quasi-TM modes.
An upper oxide cladding layer of 6.5 \textmugreek m covers (and fills) the waveguide structures. 
The grating structure itself is produced with a single etch fully through the waveguide layer.
In the results, we analyze the effects of lithographic feature size limits.

\subsection{2D simulation: sweep grating structure parameters}\label{sec:2Dsim}
In a given material stack, we calculate $\alpha$, $\theta$, and $n_\mathrm{eff}$ as a function of $\Lambda$ and $DC$ via 2D FDTD or finite element simulations.\footnote{If the same material stack will be used for devices at different wavelengths, simulation time could be greatly reduced by running the sweep with FDTD simulations, as the results for all wavelengths of interest can be extracted from the same individual simulation.}
For each combination of $\Lambda$ and $DC$, we fit the emitted field profile to find $\alpha$ and $\theta$.
Extending beyond the approach of \cite{Mehta_Ram_2017}, we use the rate of phase accumulation within the grating to find $n_\mathrm{eff}$.

\begin{figure}[!t]
\centering
\includegraphics[width=3.3in]{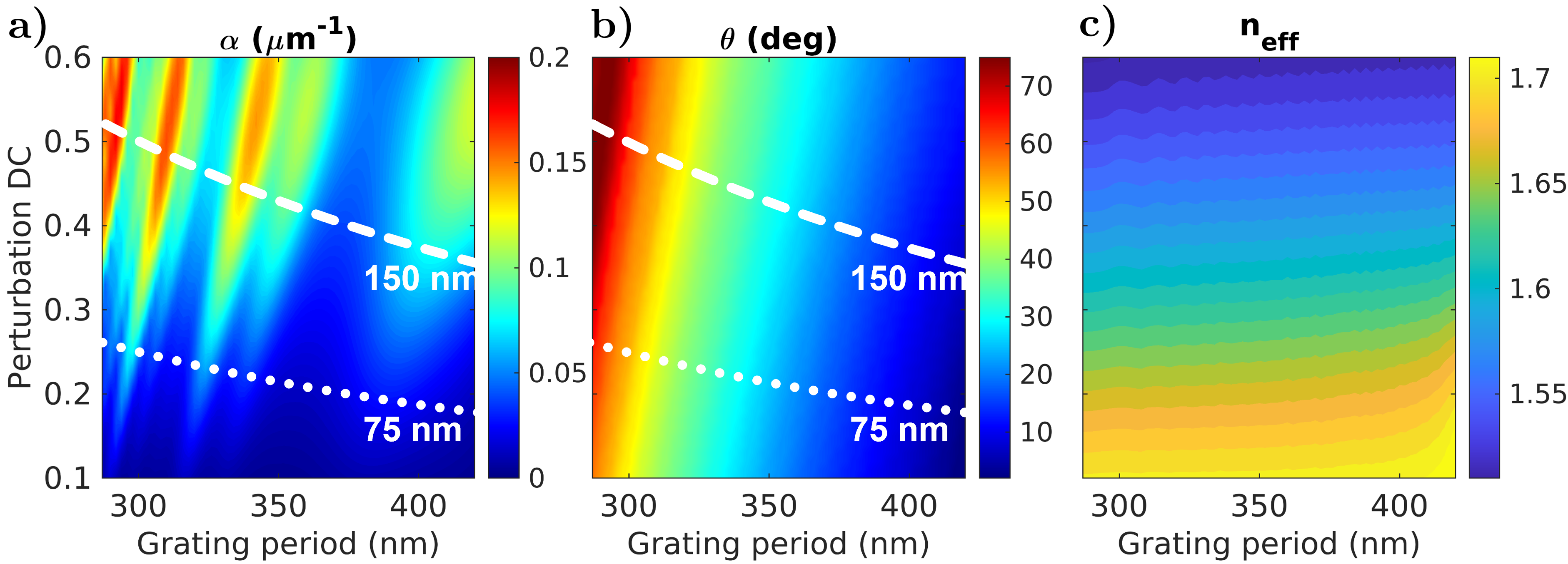}
\caption{Lookup tables resulting from 2D simulations for 732 nm light in Si$_3$N$_4$.
Emission angle corresponds to the angle in vacuum above the oxide cladding.
This regime is backward-emission ($\theta$ defined as the angle from normal, positive in the direction of the input waveguide).
The thresholds for minimum feature sizes of 75 and 150 nm are indicated.
The slightly tilted contours of (c) indicate that the duty cycle alone cannot describe $n_\text{eff}$, despite the simple structure of a fully etched waveguide.
}
\label{fig:LUTs}
\end{figure}

Figure \ref{fig:LUTs} shows the resulting 2D lookup tables for the 732 nm light in the Si$_3$N$_4$ grating.
At 732 nm, we work in the regime where the first diffraction order is emitted backward, ensuring that no higher diffraction orders can exist and all the upward-scattered light is into our mode of interest.
Minimum feature size limitations significantly restrict the accessible parameter range, so for shorter wavelengths we work in the forward-emitting regime where additional diffraction orders are present.

The grating strengths calculated in Fig.~\ref{fig:LUTs}a include the effects of multiple reflections off both the substrate and oxide-vacuum interface.
Strong fringes arise from interference with the primary back-reflection from the substrate, as well as weaker higher-order fringes which, based on their periodicity and strength, stem from interference with the initially upward-diffracted beam after reflection off the vacuum interface and again the bottom substrate.
Equipped with the foundational data, we now proceed to design individual gratings for specific beam profiles.

\subsection{Reverse propagation of desired emission}\label{sec:back-prop}
Design of a specific device begins with the desired emitted beam and its corresponding field in the grating, calculated from the target field's propagation backward toward the chip.
We then use (1) the field at the chip-vacuum interface to extract the intensity and local emission angle profiles to produce the grating structure apodization via the lookup table data, and (2) the field in the waveguide plane to determine the curvature of the grating lines through holographic interference with the incident guided field in the grating region.
For analytical propagation of Gaussian modes in the paraxial approximation, we utilize the complex $q$-parameter method presented in \cite{Kochkina_2013}, maintaining generality for astigmatic and elliptical focusing, and accounting for refraction at oblique incidence.

\begin{figure}[!t]
\centering
\includegraphics[width=3in]{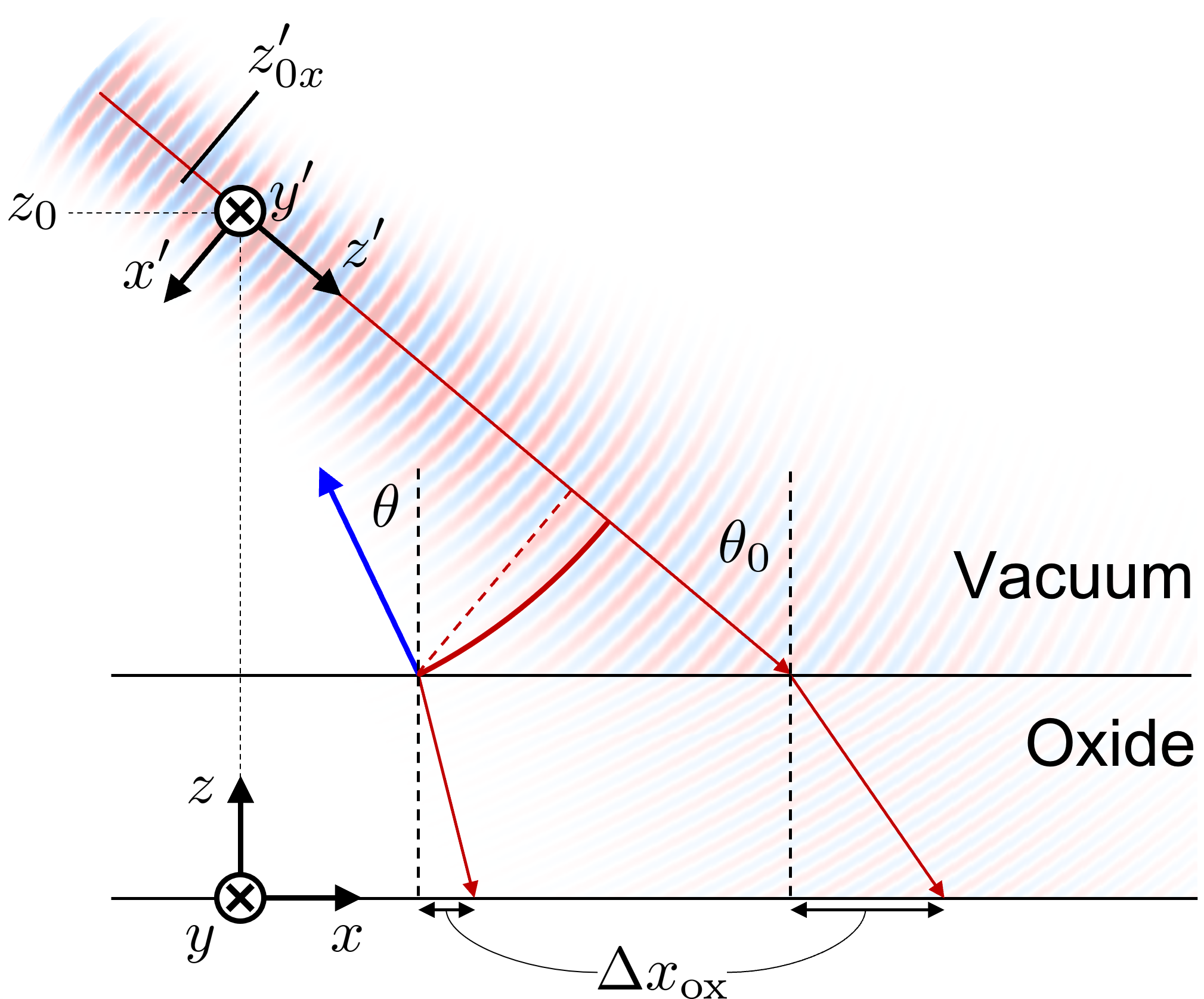}
\caption{Coordinate frame for reverse propagation of the desired beam toward the chip. 
We set the origin to the ion position ($z=z_0$ in the waveguide frame) and share the $y$ axis with the waveguide frame, with the beam axis $z'$ directed toward the chip at $\theta_0$ from normal.
The focal points of each axis, $z_{0x/y}'$, can be varied independently. 
The resulting field at the oxide surface provides the desired grating strength and emission angles, and the phase in the waveguide plane ($z=0$) is interfered with that in the waveguide to extract the grating lines curvatures.}
\label{fig:beam_coordinates}
\end{figure}

As shown in Fig.~\ref{fig:beam_coordinates}, we define $z'$ as the beam propagation axis toward the chip, and set the beam-frame $y'$-axis equal to that of the chip-frame.
We define distinct focal waists $w_{0x}$ and $w_{0y}$ along the $x'$ and $y'$ axes and allow their focal points along $z'$ to be at independent positions ($z_{0x}'$ and $z_{0y}'$). 
The orthogonality of $x'$ and $y'$ allows for the treatment of each component independently \cite{Kochkina_2013}.
In the following, we describe the propagation calculation which is performed for each axis separately (omitting an axis subscript for notational clarity).

For each axis, we define 
\begin{equation}\label{eq:qparam}
    q(z'-z_0') = z'-z_0' + i z_{\text{R}},
\end{equation}
where $z'-z_0'$ is the distance from the focal point, and $z_{\text{R}} = \frac{\pi w_{0}^2}{\lambda}$ the Rayleigh range.
Alternatively, this can be expressed as
\begin{equation}\label{eq:invq}
    q^{-1} = \frac{1}{R} - i \frac{\lambda}{\pi w^2},
\end{equation}
showing the connection to the phase front radius of curvature $R$ and the beam waist $w$.

The $q$-parameter representation allows simple description of beam propagation.
The changes induced by any particular element are described by its \textit{ABCD} matrix, $\big(\begin{smallmatrix} A & B\\ C & D \end{smallmatrix}\big)$, the elements of which transform the beam parameter as \cite{alda2003laser}
\begin{equation}\label{eq:qinv_ABCD}
    q_2^{-1} = \frac{C + D q_1^{-1}}{A + B q_1^{-1}}.
\end{equation}
Rather than tracking distance from a focal point, propagation is simply described via application of this transformation.
To reflect this, we drop notation implying explicit functions of $z'$.

The relevant beam parameters at any point can be extracted from \eqref{eq:invq}, allowing the electric field to be fully specified by the $q$-parameters:
\begin{equation}\label{eq:Efield_qparams}
    \Vec{E}(x',y',z') = \Vec{E_0}(z') e^{-i\left[ k\left( \frac{x'^2}{2q_x} + \frac{y'^2}{2q_y} \right) + \phi_\text{ac} - \eta\right]}
\end{equation}
where $k=2\pi/\lambda$ and
\begin{equation}\label{eq:E0vec}
    \Vec{E_0}(z') = \Vec{\epsilon}\sqrt{\frac{P k \sqrt{z_{R,x}z_{R,y}}}{\pi |q_x q_y|}}
\end{equation}
includes the polarization vector $\Vec{\epsilon}$ and the total power in the beam $P$.
The sum of the propagated optical path lengths through each media provides the plane wave phase accumulation $\phi_\text{ac}=k_0\sum_i n_i d_i$, and the Gouy phase is given as
\begin{equation}\label{eq:Gouy}
    \eta(z') = \frac{1}{2}\left[ \text{arctan}\left( \frac{\operatorname{Re}(q_x)}{\operatorname{Im}(q_x)}\right) + \text{arctan}\left( \frac{\operatorname{Re}(q_y)}{\operatorname{Im}(q_y)} \right) \right]. 
\end{equation}



In transforming the $q$-parameters, propagation a distance $d$ through a medium of index $n$ is described by the matrix
\begin{equation}\label{eq:ABCD_prop}
    \begin{pmatrix}
    1 & d/n \\
    0 & 1
    \end{pmatrix}, 
\end{equation}
and the component matrices for refraction of a beam incident at $\theta$ are
\begin{center}
    \begin{tabular}{cc cc}
        \multicolumn{2}{c}{\underline{Sagittal ($y'$)}} \quad & \multicolumn{2}{c}{\underline{Tangential ($x'$)}} \\ [1.0ex]
        \multicolumn{2}{c}{$\begin{pmatrix}1 & 0 \\ 0 & \frac{1}{n_{r}}\end{pmatrix}$} \quad &
        \multicolumn{2}{c}{$\begin{pmatrix}\frac{\sqrt{n_{r}^2 -\text{sin}^2\theta}}{n_{r} \text{cos}\theta} & 0 \\ 0 & \frac{\text{cos}\theta}{\sqrt{n_{r}^2 -\text{sin}^2\theta}}\end{pmatrix}$} \\
    \end{tabular}
\end{center}
\noindent where $n_{r} = n_2/n_1$ is the ratio of output to input indices (in our case, $n_r = n_\text{oxide}/n_\text{vac}$) \cite{Kochkina_2013}.
These are applied to the $q_y$ and $q_x$ parameters, respectively.
Note that both radii of curvature $R_x$ and $R_y$ change upon refraction, but only the beam waist $w_x$ is affected.

After transforming the oxide-vacuum interface into beam-frame coordinates $(x', y', z')$, at each $z'$ we calculate $q_x$ and $q_y$ (using the propagation distance from their focal points, $z' - z_{0x/y}'$), and obtain the full field from (\ref{eq:Efield_qparams}).
Because we only consider TE polarization ($\vec{E} \parallel \hat{y}$), no transformation between field components is required. 

Where the beam axis is incident on the chip surface we calculate the central $q_{x,1}$ and $q_{y,1}$ and apply the refractive transformations to arrive at post-refraction $q_{x,2}$ and $q_{y,2}$ values. 
The $q$-parameters at any point in the oxide are now defined by their propagation distance from this central point along the new in-oxide beam axis (its angle redefined using Snell's Law), so applying (\ref{eq:ABCD_prop}) with $n=n_\text{oxide}$ we similarly obtain the full electric field at each point in the waveguide plane, whose phase we define as $\phi_\text{em}$, for use in the transverse design described in Sec. \ref{sec:transverse}.

As we proceed to specifying the grating in the waveguide plane with data that directly maps grating geometry to the emission \emph{in vacuum}, note that we also account for the lateral offsets due to propagation through the upper oxide ($\Delta x_\text{ox}$ in Fig.~\ref{fig:beam_coordinates}) by using the desired beam's instantaneous angles of incidence.

\subsection{Longitudinal design: mapping emission to structural parameters}

\begin{figure}[!t]
\centering
\includegraphics[width=3.3in]{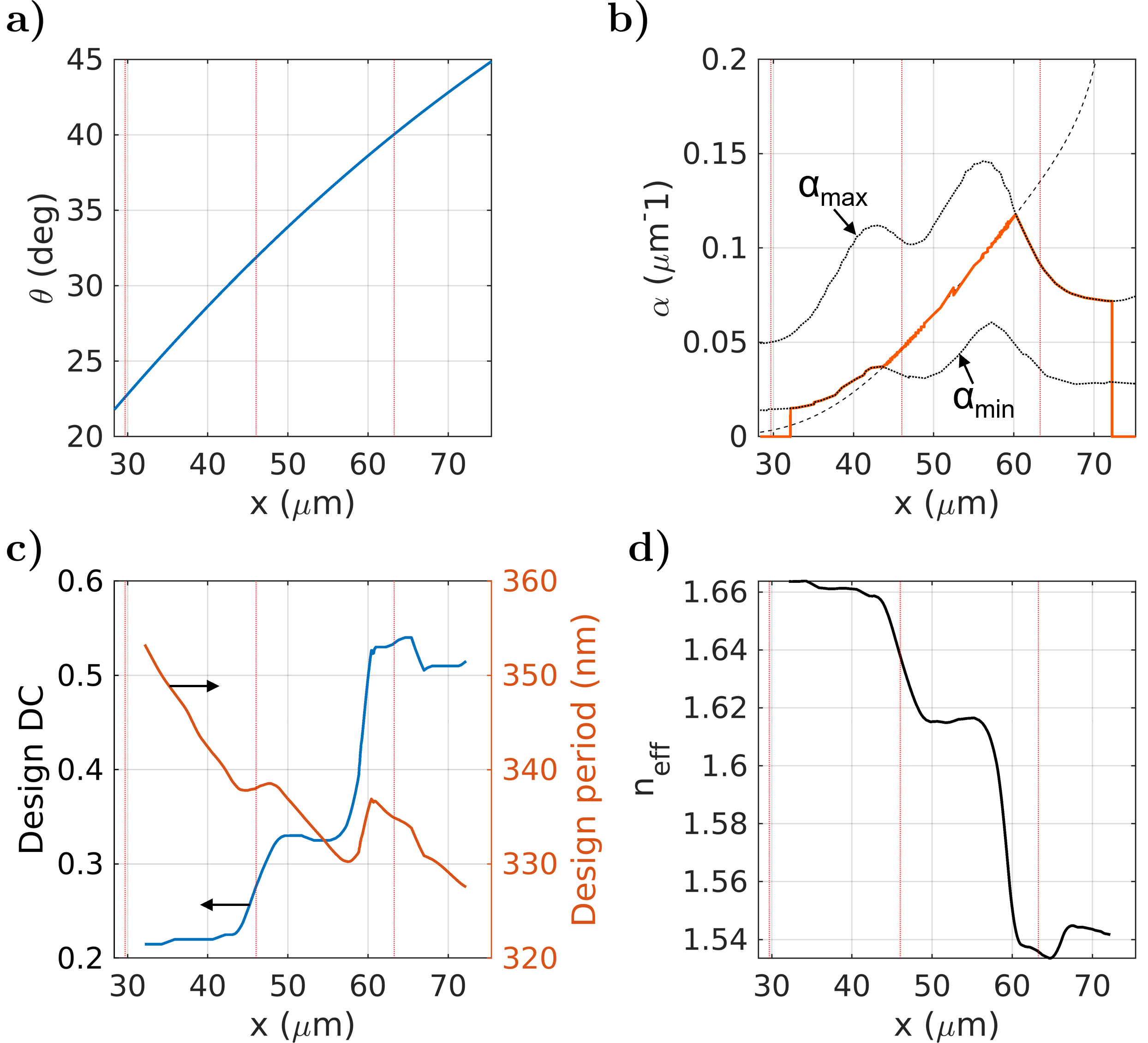}
\caption{Longitudinal emission and grating structure. 
The local angle of emission (a) and desired grating strength (b, dashed) for a 732 nm beam at 32\degree, focusing 70 \textmugreek m above the surface to $w_{0x}=w_{0y}=1.5$ \textmugreek m.
To keep longitudinal parameters invariant with respect to the taper start position, the $x$-position shown is relative to the ion.
The lower threshold for $\alpha$ is set by a minimum feature size of 75 nm (dotted line in Fig.~\ref{fig:LUTs}).
Selecting at each $\theta$ the nearest possible $\alpha$ (b, orange) provides the resulting period and duty cycle (c) which also results in a varied effective index in the grating region (d).
Red dotted lines indicate where the beam axis and $1/e^2$ intensity levels (corresponding to the beam waists $w_x$) intersect the waveguide plane.
Truncation parameters $\chi_1=0.85$ and $\chi_2=1.52$ define the start and end of the grating relative to these positions.
}
\label{fig:DesignParams}
\end{figure}

Control of the emitted beam along the longitudinal direction comprises tuning both the instantaneous emission angle $\theta(x)$ and the grating strength $\alpha(x)$ along the length of the grating.
We define the emission angle as the angle between the the chip normal and the surface normal of the Gaussian phase front (Fig.~\ref{fig:beam_coordinates}).
This can be calculated analytically from expressions for the phase-front radii of curvature or through numerical integration.
Figure \ref{fig:DesignParams}a shows the resulting $\theta_\text{em}(x)$ for a 732 nm beam at $32\degree$ focusing to a 1.5 \textmugreek m spot.

The required $\alpha(x)$ is calculated from the normalized field intensity $|E(x)|^2$ using 
\begin{equation}
    \label{eq:alpha-integral}
    \alpha(x) = \frac{\eta \, |E(x)|^2}{2\left(1-\eta\; \int_{x_0}^{x}|E(x')|^2\; \text{d}x'\right)},
\end{equation}
where $x_0$ is the start of the grating, and $\eta$ the desired fraction of power outcoupled by the end of the grating \cite{Zhao_Fan_2020} (we target $\eta=1$ in current designs).
For our example beam, $\alpha(x)$ is shown in Fig.~\ref{fig:DesignParams}b, where the influence of material and fabrication constraints is evident in $\alpha_\text{min}$ and $\alpha_\text{max}$ which arise from the minimum etched feature size and the material stack, respectively.

Such limitations hamper control of the intensity profile.
For example, the small initial values of $\alpha$ needed for Gaussian profiles are often unattainable, leading to ``front-heavy'' emission profiles (e.g. grating \rm{I} in Fig.~\ref{fig:732_results}a below), which we mitigate by truncating the start of the grating closer to the beam center.
To maintain generality across beam parameters, we specify the grating start/end positions in reference to the desired beam waist in the waveguide plane, using $\chi_1$ and $\chi_2$ to define the fraction of the waist (distance from beam center to $1/e^2$ intensity on either side) at which we truncate the grating.

Obtaining the grating structures entails mapping each pair $\theta(x)$ and $\alpha(x)$ to a unique period and duty cycle pairing, $\Lambda(x)$ and $DC(x)$.
Being especially critical for a focusing beam, our implementation prioritizes the local angle of emission, selecting the closest possible $\alpha$ for any given $\theta$ (Fig.~\ref{fig:DesignParams}b).
The resulting $\Lambda(x)$ and $DC(x)$ (Fig.~\ref{fig:DesignParams}c) corresponds to a single line traversing the contours in Fig.~\ref{fig:LUTs} while constrained by $\alpha_\text{min/max}$.
From these $(\Lambda, DC)$ pairs, the $n_\text{eff}$ contour yields the resulting $n_\text{eff}(x)$ (Fig.~\ref{fig:DesignParams}d) which will be a utilized later on to track the field's phase in the grating region. 
It would be straightforward to extend this same method to more complex grating structures such as those including multiple layers or etch depths \cite{Shirao_Klawson_Mouradian_Wu_2022}.

Material and fabrication limitations can lead to asymmetric, highly non-Gaussian intensity profiles with propagation difficult to analytically predict.
Before proceeding to the transverse structure, it is often worthwhile to verify the longitudinal design with explicit 2D simulation.


 
\subsection{Transverse design: interference in waveguide plane}\label{sec:transverse}

Diffracted emission relies on phase matching between the propagating mode in the waveguide and the resulting scattered field.
When defining the 2D longitudinal grating structure, this phase matching was implicit in the relationship between effective index, period, and emission angle. 
To produce the full 3D grating, we extrude that 2D structure along the contour lines which ensure phase matching to the desired output beam. 
In other words, the grating lines are curved according to where the two fields constructively interfere---i.e., where $\phi_\text{em} + \phi_\text{wg}$ is an integer multiple of $2\pi$, with $\phi_\text{wg}(x,y)$ the phase of the field expanding through the waveguide taper and $\phi_\text{em}(x,y)$ already calculated in Sec.~\ref{sec:back-prop}.

To calculate $\phi_\text{wg}$ in the common case of non-adiabatic tapers, we assume circular expansion of phase fronts from the taper start and integrate the optical path length from there to determine the accumulated phase at all points.
The integration can be split into two regions: the first consisting of the unperturbed taper where we use the normal slab waveguide effective index $n_\text{wg}$, and the second being the grating region, where we use the simulated $n_\text{eff}(x)$ values corresponding to the specific local grating parameters (Fig.~\ref{fig:DesignParams}d).

The integral for a given point $(x,y)$ in the taper takes the form
\begin{equation}
    \label{eq:phiwg-integral}
    \phi_\text{wg}(\rho) = k_0 \int_{0}^{\rho}n(\rho')\; \text{d}\rho',
\end{equation}
with $k_0=2\pi/\lambda$ and $\rho=\sqrt{(x-x_\text{t})^2+y^2}$ the distance from the taper start ($x=x_\text{t}$, $y=0$).
Letting $x_1$ and $x_2$ be the start and end positions of the grating, we express
\begin{equation}
n(\rho) =
    \begin{cases}
        n_\text{wg}\qquad  & \rho<x_1-x_\text{t}\;\text{  or  }\rho>x_2-x_\text{t}\\
        n_\text{eff}(\rho + x_\text{t}) & \text{otherwise},\\
    \end{cases}
\end{equation}
where $n_\text{eff}(\rho + x_\text{t})$ is used to correspond directly to the function over $x$ values relative to the ion (as shown in Fig.~\ref{fig:DesignParams}d).
The precise evolution of $n_\text{wg}$ before the grating region is not critical as it produces only a constant offset which does not influence the interference calculation.

We now turn to the task of extracting the interference contours. 
For complex phase profiles such as Laguerre-Gaussian beams, we extract these grating lines explicitly as vectors of points.
When coupling between the fundamental waveguide and free-space modes, however, the lines retain a symmetry about the central $x$ axis and can be expressed as even polynomials of $y$ whose coefficients are functions of the longitudinal position $x$.

We find that the inclusion of a quartic term, beyond the parabolic profiles often assumed \cite{Mehta_Ram_2017, Khan_Combrie_DeRossi_2020}, is required for accurate performance in many cases---especially in forward-emitting gratings, where the phase fronts of the emitted beam curve in the opposite direction as those expanding within the taper. 
Thus, we use grating lines which take the form 
\begin{equation}
    \label{eq:quartic-polynom}
    x = x_i + \frac{1}{2R} y^2 + A y^4,
\end{equation}
where $x_i$ is the origin point on the centerline, $R$ the radius of curvature \cite{Mehta_Ram_2017}, and $A$ the quartic coefficient.

\begin{figure}[!t]
\centering
\includegraphics[width=3.3in]{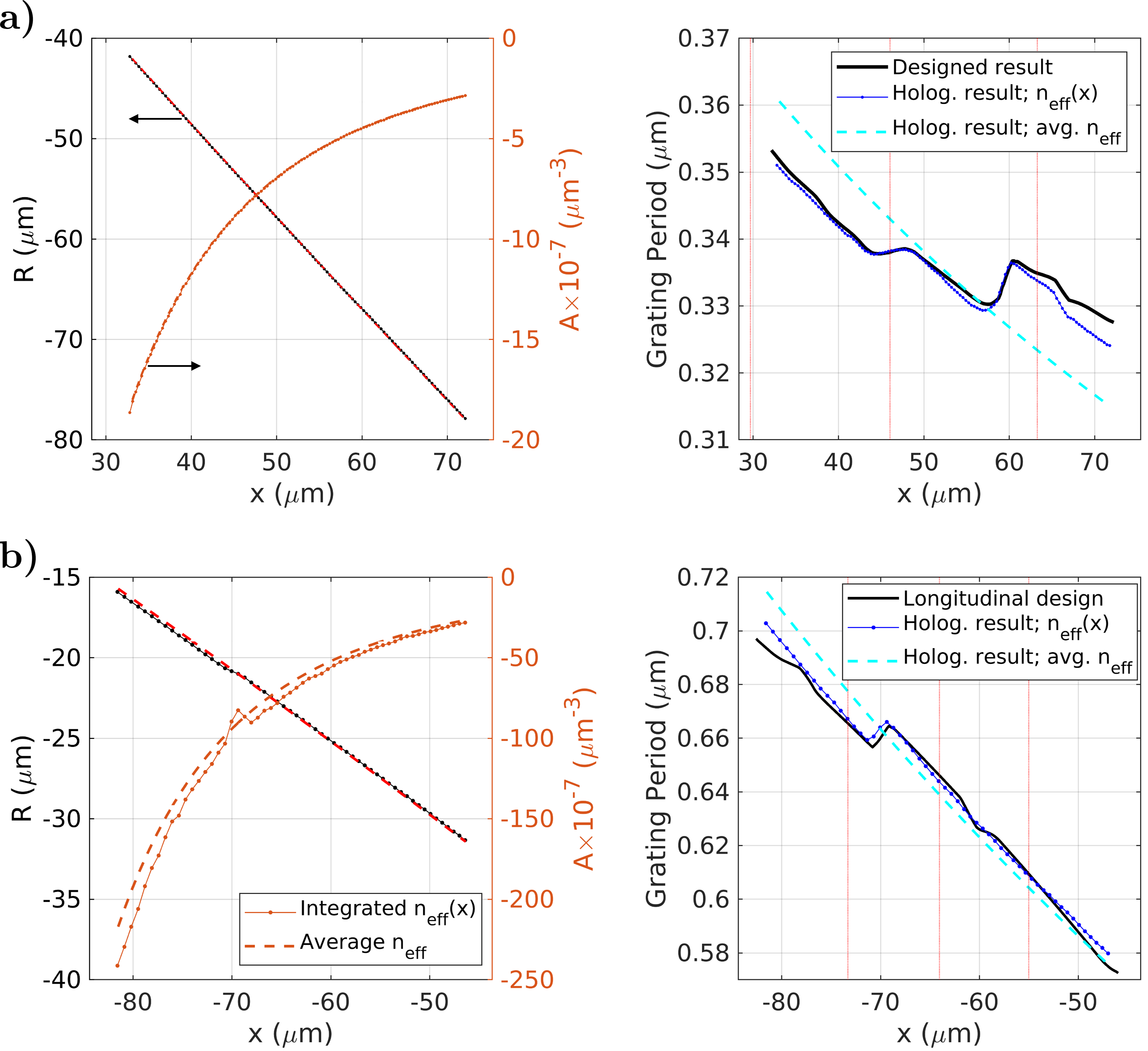}
\caption{Outputs from holographic interference contours. 
Polynomial coefficients for grating line curvatures (left) and the corresponding periods in agreement with the designed values (right). 
Dashed lines show the results if the average value of $n_\text{eff}(x)$ is used instead of integrating.
Designs shown are (a) the 732 nm beam from Fig.~\ref{fig:DesignParams} and (b) a forward-focusing 532 nm beam for comparison ($\theta_0=-41$\degree, $z_0=76.5$ \textmugreek m, $w_{0x}=w_{0y}=2.5$ \textmugreek m).
}
\label{fig:output_coeffs}
\end{figure}
Extracted coefficients and the resulting periods are shown in Fig.~\ref{fig:output_coeffs} for the 732 nm backward-focusing and a 532 nm forward-focusing beam for comparison. 
The quartic coefficients in the forward-focusing case are over an order of magnitude larger than the backward focusing, and in designs with the taper start closer to the grating these were multiple times larger still.
We also observe a substantial change when including the integration over $n_\text{eff}(x)$ (orange solid vs. dashed in Fig.~\ref{fig:output_coeffs}b).

The distance in $x$ between the extracted contours also describes grating periods (blue, solid in Fig.~\ref{fig:output_coeffs}). 
Their agreement with the designed periods supports our treatment of $n_\text{eff}(x)$ with an assurance of self-consistency.

\subsection{Taper angle and accounting for focal shift}

As a first step toward determining the taper angle of our devices, we adopt the approach of \cite{Mehta_Ram_2017}, optimizing the transverse 1D overlap between the waveguide mode and the desired emission at the central point, where the emitted field intensity is designed to be greatest (and the beam waist widest).
For the fundamental Gaussian mode and the approximate $\text{cos}^2$ profile in the waveguide, this corresponds to setting the grating width at that point to be $2.844 w_y$, where $w_y$ is the transverse beam waist in the waveguide plane \cite{Mehta_Ram_2017, Mehta_2017}.
Taper angles prescribed using this method largely agree with those from the full 2D overlap integral (calculated in \cite{Zhao_Fan_2020}).

Even with the refinements developed above, the resulting devices produced transverse focuses shifted relative to their designed position along the beam axis, consistently toward the chip by distances from one half to a full Rayleigh range. 
These proved sensitive to the taper angle, reducing in scale as the taper angle, and thus grating width, is increased.

Precisely positioned focuses necessitate understanding and accounting for these shifts, and investigations revealed that they can be predicted accurately using the formulation of \cite{Li_Wolf_1982} for truncated focused Gaussian beams.
 
We first review the general theory and then proceed to detail its extension to grating outcouplers. 
We consider a Gaussian beam focused by a thin lens with focal length $f$ and a circular aperture of radius $a$. 
Given a beam waist $w$ at the aperture, the truncation parameter is defined as $\xi=(a/w)^2$
We define the Fresnel number as 
\begin{equation}
    N = \frac{2}{\lambda}(f-\sqrt{f^2 - a^2}),
\end{equation}
adopting the form which is valid beyond the paraxial regime \cite{Sheppard_Torok_1998}.
Then, we define the dimensionless parameter 
\begin{equation}
    \label{eq:u}
    u=\pi N \frac{z}{f+z}
\end{equation} where $z$ is the location along the beam axis relative to the geometric focus.
Locations of intensity extrema are provided by the (potentially many) roots of the transcendental equation
\begin{equation}
    \label{eq:transcendental}
    \frac{\left(u + \frac{\xi^2}{\pi N}\right)\left(\text{cosh}(\xi) - \text{cos}(u)\right)}{\left(1-\frac{u}{\pi N}\right)\left(\xi^2 + u^2\right)} = \text{sin}(u), 
\end{equation}
with the true peak intensity corresponding to the root in the range $(-2\pi,\, 0)$ \cite{Li_Wolf_1981}.
After solving numerically, we use \eqref{eq:u} to convert back to the true focal shift which will always be negative (toward the lens). 

To apply this theory to our grating outcoupler, we must identify reasonable analogs to the lens and aperture system.
We consider the beam only in vacuum, treating the oxide as part of the ``lens.''
For the focal length, we use the distance from the focus position $z_{0y}'$ to the chip surface along the beam axis. 
For the aperture, we first take into account the potentially imbalanced intensity profile resulting from our designed $\alpha(x)$ values (\cite{Zhao_Fan_2020}), 
\begin{equation}
    \label{eq:intensityfromalpha}
    I(x) = 2\alpha(x) \exp{\left[-2 \int_0^x \alpha(t) dt\right]}, 
\end{equation}
and consider the $x_\text{mean}$ position corresponding to the distribution's center of mass.
In the absence of a top oxide, we would define the aperture radius as the $y$ value where the circular arc from the taper start to $x_\text{mean}$ intersects the taper edges.
To obtain the corresponding value at the oxide-vacuum interface, we further account for the lateral offsets of propagation through the oxide (e.g. $\Delta x_\text{ox}$ in Fig.~\ref{fig:beam_coordinates}).
Adjusting the taper angle effectively adjusts this aperture, tuning the focal shift accordingly.

Correcting for transverse focal shift is done in two stages: (1) an initial target offset $z_{0y}'$ at the outset of design, and (2) adjustments to the taper angle at this final stage.
Specifically, we first determine the offset $z_{0y}'$ using an aperture radius of $1.422 w_y$ at the chip surface (calculated using the basic $w_y = w_{0y}\sqrt{1 + (z'/z_{\text{R}y})^2}$ expression).
With the desired beam parameters then settled, we proceed with the design process, producing the $\alpha(x)$ required to calculate $x_\text{mean}$.
From this, we calculate more precisely the resulting shift as a function of the aperture width and finalize the taper angle.
Throughout the results below, we use $\Delta\gamma$ to describe the expansion of the taper angle relative to the $\gamma_0$ initially prescribed.

\section{Results}

In this section we present a selection of gratings produced with the above design process and their resulting emission according to 3D FDTD simulation.

We first verify the focal shift theory presented above by examining the relationship between taper angle and focal shift in an ideal material stack to avoid confounding influences.
Proceeding to outcouplers using the full material stack, we present results including micron-scale backward- and forward-focusing and examine the influence of various design considerations such as fabrication limits, emission angles in the presence of substrate back-reflection, and truncation parameters.
We then describe and demonstrate the generation of Hermite-Gaussian modes and tightly focused Laguerre-Gaussian vortex beams of different orders.

\subsection{Focal shift and non-Gaussian profiles}
\begin{figure}[!t]
\centering
\includegraphics[width=3.3in]{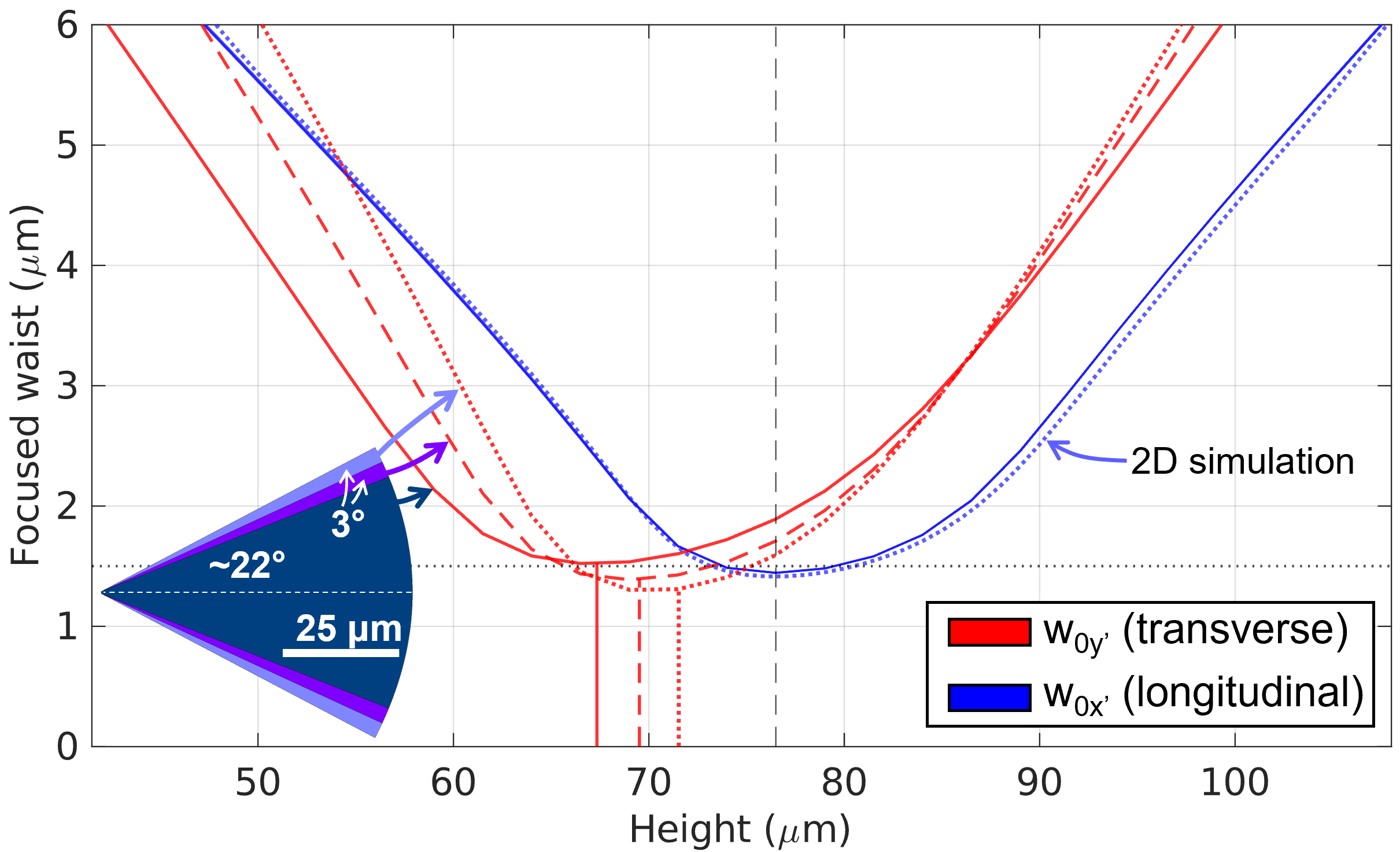}
\caption{Predicting focal shifts.
Transverse (red) and longitudinal (blue) beam waists are shown for three outcoupler variants with increasing taper angle.
In the transverse focuses we observe shifts from the designed focal height (76.5 \textmugreek m), with the deviation being reduced for increasing taper angle (red solid, dashed, and dotted lines correspond to taper angles of $\gamma \approx 22\degree + \Delta \gamma$ for $\Delta \gamma = 0\degree,3\degree,6\degree$).
The longitudinal waists for all three variants are plotted but visually indistinguishable.
Predicted transverse focal positions are indicated as vertical red lines, and longitudinal emission from 2D simulation (blue, dotted) also shows strong agreement with the final results.
With 732 nm light, the Si$_3$N$_4$ outcouplers were designed to focus emission at $\theta_0=25\degree$ to 1.5 \textmugreek m waists.
Design/simulation included no silicon substrate and a minimum feature size of 36 nm.
Longitudinal truncation parameters set to $\chi_1=\chi_2=1.52$, the taper start $x_\text{t} = -10$ \textmugreek m.
Waists are extracted from Gaussian fits at slices of constant height, with the longitudinal values corrected by a factor of $\text{cos}(\theta_0)$ to correspond to the $x'$ axis orthogonal to the beam.
}
\label{fig:732_noSi}
\end{figure}

For simplicity in investigating the focal shift's relationship to taper angle, we designed devices for an altered material stack (1) omitting the back-reflecting substrate and (2) including minimal fabrication limitations ($DC>0.1$, or a minimum feature size of 36 nm).
An outcoupler was designed in Si$_3$N$_4$ to focus 732 nm light emitted at $\theta_0=25\degree$ to $w_{0x'}=w_{0y'}=1.5$ \textmugreek m at a height of $z_0=76.5$ \textmugreek m (70 \textmugreek m above the surface).
The designed focal position includes no offsets ($z_{0x/y}'=0$), and the initial taper angle is set to $\gamma_0\approx22\degree$.
In Fig.~\ref{fig:732_noSi} we compare the resulting emission from this outcoupler to that from variants with the taper further expanded by $3\degree$ and $6\degree$.

The results confirm that the theory of the previous section accurately predicts both the transverse focal shift and its variation with the taper angle, decreasing as the effective transverse aperture expands (Fig.~\ref{fig:732_noSi}).
Increasing the taper angle had negligible impact on the overall efficiency of the devices, while slight distortions in transverse profile explain the narrowing observed in the focused waists in alignment with expectations from such non-Gaussian profiles \cite{Parent_Morin_Lavigne_1992}.
Profile distortions limit the extent to which taper angle can be increased to mitigate focal shifts, so our designs typically combine few-degree expansions of the taper angle with built-in offsets $z_{0y}'$.

Although the longitudinal focus occurs at precisely the designed height here, this does not imply that the longitudinal direction is immune to aperture-related focal shifts (we observe this with differing truncation parameters in Fig.~\ref{fig:467_results_fund}a below).
In this case, the distortions introduced by the 36 nm minimum feature size coincidentally counter the few-micron shift which would be otherwise expected, highlighting how simple analytical models are unsuited to the often asymmetric and irregular longitudinal field profiles.
In contrast, strong agreement with the 3D simulation results demonstrates the accuracy of 2D numerical simulation in predicting the longitudinal propagation.

\subsection{Tight focusing}\label{sec:tightfocusing}

Micron-scale focusing gratings were designed at various wavelengths and emission angles to enable individual ion addressing with high peak intensities.
Here, we present both backward- and forward-focusing gratings, in the material stack of Fig.~\ref{fig:material_stack}, at different wavelengths, waists, and focal heights.
Variations of each are used to investigate the influences of fabrication limits, substrate back-reflection, longitudinal truncation parameters, and taper angle.

\subsubsection{732 nm, 1.5 \textmugreek m focus}

\begin{figure}[!t]
\centering
\includegraphics[width=3.3in]{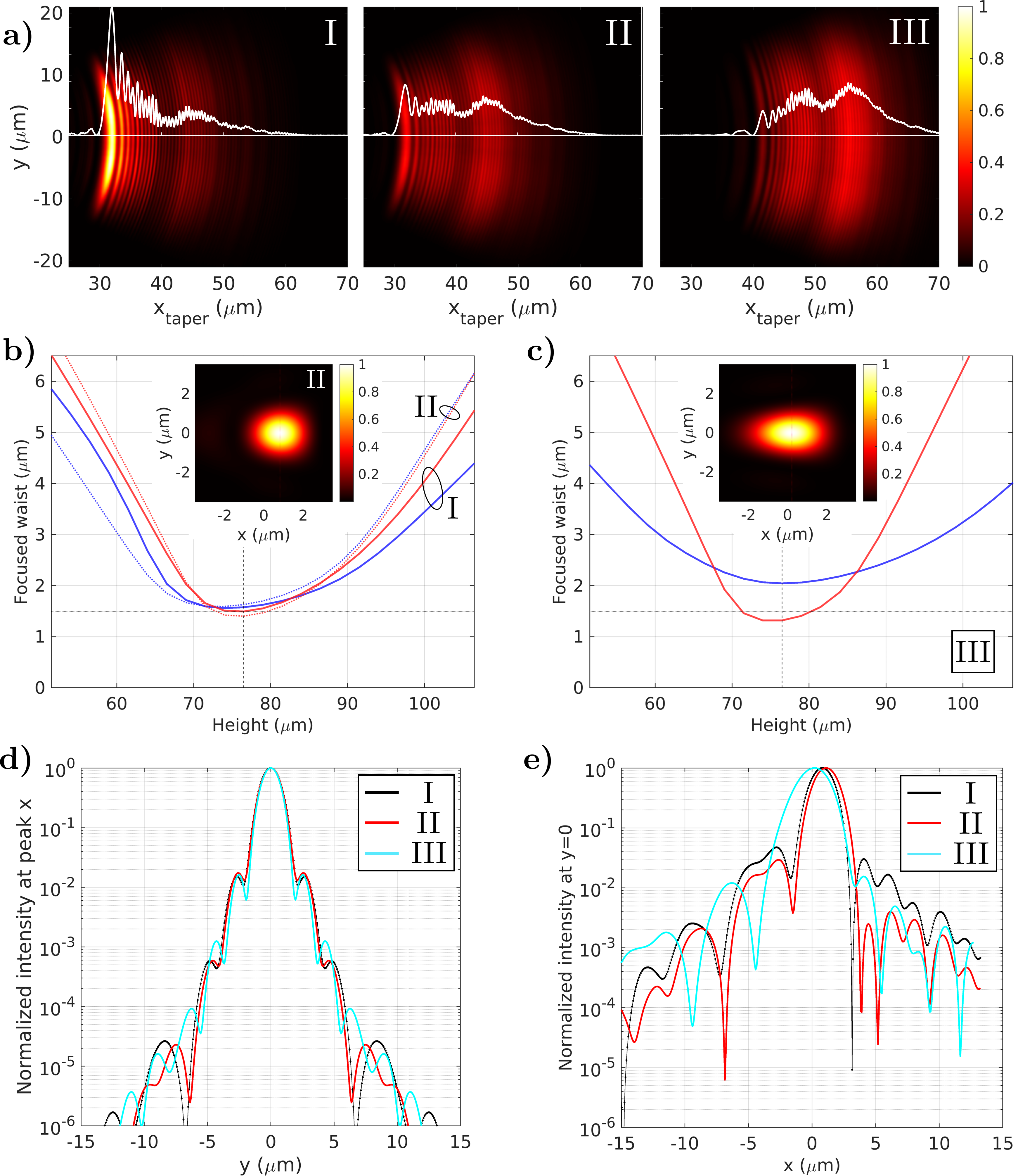}
\caption{Comparison of three tightly-focusing, back-emitting 732 nm outcouplers, designed for $w_{0x}=w_{0y}=1.5$ \textmugreek m at $z_0=76.5$ \textmugreek m.
(a) Simulated intensities 0.5 \textmugreek m above the oxide surface with slices along the centerline overlain. 
Values normalized to the peak in \rm{I}.
Gratings \rm{I} and \rm{II} both emit at $\theta_0=25\degree$, but have minimum feature sizes of 150 nm and 75 nm, respectively (corresp. to the lines in Fig.~\ref{fig:LUTs}). Grating \rm{III} shares the 75 nm limit but emits at $\theta_0 = 32\degree$.
(b, c) Resulting longitudinal (blue) and transverse (red) beam waists vs. height. Insets show horizontal slices of intensity in the ion plane ($z = 76.5$ \textmugreek m) for \rm{II} and \rm{III}.
(d, e) Intensity cross sections in the ion plane: (d) transverse, at peak intensity; (e) longitudinal, along the centerline. Gratings \rm{I}, \rm{II}, \rm{III} shown in black, red, cyan, respectively.
All variants were truncated at $\chi_1=0.85$, $\chi_2=1.52$.
Correcting for focal shift, they were designed with $z_{x,y}'=5$ \textmugreek m, and taper angles expanded by $\Delta \gamma = 4\degree$, resulting in net taper angles of approximately 23$\degree$ (\rm{I}, \rm{II}) and 20$\degree$ (\rm{III}).
}
\label{fig:732_results}
\end{figure}

Motivated by \cite{Beck_2020}, back-emitting Si$_3$N$_4$ outcouplers were designed for 732 nm light, targeting 1.5 \textmugreek m waists 70 \textmugreek m above the chip.
Their simulated emission is compared in Fig.~\ref{fig:732_results}.
Devices \rm{I} and \rm{II} both emit at $\theta_0=25\degree$ but incorporate minimum feature sizes of 150 nm and 75 nm, respectively.
This restricts outcoupler \rm{I} to higher grating strengths for the same local emission angles, producing the strong peak of initial intensity (Fig.~\ref{fig:732_results}a) and ultimately leading to more pronounced longitudinal sidelobes at the focal plane (Fig.~\ref{fig:732_results}e, black vs. red) and slightly increased asymmetry of the beam waists about the focal height (Fig.~\ref{fig:732_results}b, solid vs. dotted).
Accompanied by a negligible reduction in efficiency ($<$0.6\%), these results demonstrate that fabrication limits have limited impact on the resulting beam, provided that the beam can still be emitted across the full emission aperture.
The asymmetric and non-Gaussian longitudinal intensity profiles of both cases lead to (1) a slight ($<1$ \textmugreek m) lateral shift of the focal point and (2) counteracting of the diffractive broadening which would otherwise be expected from truncation at $\chi_1=0.85$ and $\chi_2=1.52$, thereby still obtaining the designed longitudinal focus of $w_{0x}=1.5$ \textmugreek m.
In comparison, the more symmetric profile from device \rm{III} demonstrates focusing with less lateral shift but broadened to $\sim2$ \textmugreek m. 
With the same truncation parameters and a minimum feature size of 75 nm, the only distinction of this device is its design at $\theta_0=32\degree$.
This places the initial $\theta(x)$ at the grating start at a local minimum of $\alpha_\text{min}$---directly in the destructive interference fringe in Fig.~\ref{fig:LUTs}a---which mitigates the initial intensity peak. 
Furthermore, the emission is centered about the angles with strongest constructive interference with substrate back-reflection, enabling substantially higher efficiency (59\%) in comparison to devices \rm{I} and \rm{II} ($\sim$48\%).
However, various figures of merit may be relevant for free space applications, and here we see the reduced efficiency of devices \rm{I} and \rm{II} nearly compensated by their tighter focusing: despite an efficiency 25\% greater, the peak focused intensity from \rm{III} is only 6\% greater than that from \rm{I} and \rm{II}.



\subsubsection{467 nm, 2 \textmugreek m focus}
\begin{figure}[!t]
\centering
\includegraphics[width=3.3in]{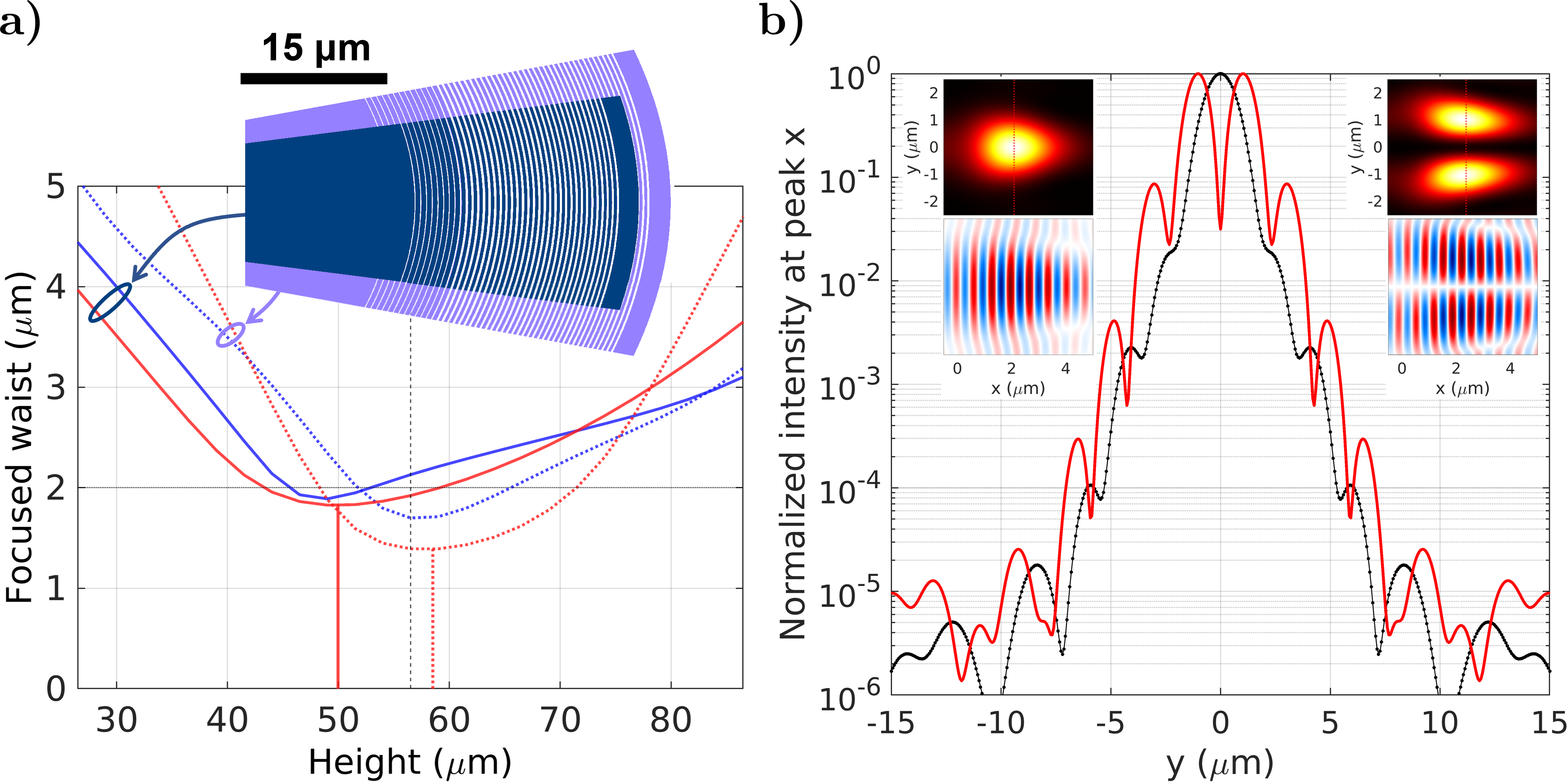}
\caption{Forward-focusing 467 nm outcouplers at $\theta_0 = -42\degree$, designed for $w_{0x}=w_{0y}=2$ \textmugreek m at $z_0=56.5$ \textmugreek m (with offsets $z'_{x,y}=10$ \textmugreek m). (a) Focused longitudinal/transverse beam waists (blue/red) of outcoupler variants shown in inset (smaller outcoupler corresponds to solid lines). Transverse focal points are accurately predicted by focal shift theory (vertical red lines) before and after expanding the taper angle by $3\degree$ (from $\gamma_0=7.4\degree$; increasing width $\sim8$ \textmugreek m from $\sim$18.5 \textmugreek m).
Extending the grating length $\sim6$ \textmugreek m via truncation parameters leads to a forward shift of the longitudinal focus. (b) Focused HG$_{10}$ beam produced using a higher order waveguide mode (TE$_1$) incident on the larger grating coupler variant of (a) (only physical difference is wider input waveguide to accommodate mode). Insets show horizontal slices of intensity and field $E_y$ at $z=56.5$ \textmugreek m for the fundamental mode (left) and higher order (right). Main plot compares the transverse intensity profiles at the peak $x$-value (sliced along the red dotted line in top insets).}
\label{fig:467_results_fund}
\end{figure}

Due to our minimum feature sizes of $\sim$150 nm, designs for $\lambda< 500$ nm were based on forward-emission from Al$_2$O$_3$ waveguides. 
For experiments aiming to drive the $S_{1/2} \leftrightarrow F_{7/2}$ octupole transition in $\mathrm{Yb}^+$ 
\cite{Furst_Yeh_Kalincev_Kulosa_Dreissen_Lange_Benkler_Huntemann_Peik_Mehlstaubler_2020}, 
we designed outcouplers to focus 467 nm emission at $\theta_0 = -42\degree$ to 2 \textmugreek m waists, 50 \textmugreek m above the chip surface (with initial offsets $z'_{x,y}=10$ \textmugreek m to account for focal shift).
We compare the results of two variants in Fig.~\ref{fig:467_results_fund}a: one with longitudinal truncation parameters $(\chi_1, \chi_2) = (1, 1.4)$ and the ``default'' taper angle of $\gamma_0 = 7.4\degree$, and another expanded version with $(\chi_1, \chi_2) = (1.3, 1.75)$ and $\Delta\gamma = 3\degree$.

Theory accurately predicts the transverse focal shift, but with the wider angle we observe an over-focusing of the waist more substantial than in backward-focusing devices (Fig.~\ref{fig:732_noSi}) despite a less-perturbed transverse intensity profile from the shallower taper, highlighting the increased sensitivity of forward-focusing.


Despite their substantial length, over 30\% of the input power continues past the end of these gratings, with efficiencies being primarily limited by the low maximum grating strength of the Al$_2$O$_3$ waveguide owing to a relatively low refractive index contrast with SiO$_2$.
Approximately 32\% is emitted into the primary diffraction order at 42$\degree$ forward, and around 5\% in the secondary order (4$\degree$ backward).
The efficiency into the primary order varies less than 1\% between any combination of the grating length and taper angle variations, but the tighter focusing enabled by the larger variant ultimately produces a peak intensity over 40\% higher than that of the smaller.

Gratings were also designed for the same focusing but at 100 \textmugreek m above the chip. Efficiencies were around 20\% greater due to the larger grating footprint, but the desired focal widths and positions were produced with the same accuracy as the devices presented.

\subsection{Hermite-Gaussian modes}\label{sec:HG}

The complex fields and gradients of higher order modes can provide interesting capabilities for addressing ions.
Hermite-Gaussian modes have rectangular symmetry about the axis of propagation \cite{Kogelnik}, and their higher orders possess distinct intensity nulls and field gradients which can, for example, be used to drive certain atomic transitions with minimized off-resonant couplings \cite{Vasquez_Mordini_Verniere_Stadler_Malinowski_Zhang_Kienzler_Mehta_Home_2022}, including for the $\mathrm{Yb}^+$ octupole transition targeted in our 467 nm designs \cite{peshkov2023excitation}.

With laterally defined waveguide structures enabling control of the guided mode, grating couplers are well-suited for emission with higher order structure in the transverse direction.
For example, consider a focused HG$_\text{10}$ (or TEM$_{10}$) mode, which consists of two field-antisymmetric lobes about a centerline null.
From a fundamental waveguide mode, phase matching to emit such a beam calls for grating lines with a $\pi$ phase-shift across the centerline—i.e., grating lines broken in the middle, with their top and bottom halves shifted from one another by precisely half a period.
Investigations showed that both the diffuse scatter from the centerline discontinuities as well as the asymmetry in the grating start positions lead to lower quality emission.

Alternatively, first producing a waveguide mode of the same higher transverse order (e.g. via passive mode-conversion \cite{Paredes_Mohammed_Villegas_Rasras_2021}) enables the same phase matching using continuous grating lines.
We opt for this approach, which allows higher mode purity and outcoupling efficiency.
Furthermore, we can directly employ the gratings designed for fundamental mode emission because the phase front radii-of-curvature of HG beams is independent of mode order \cite{Kogelnik} and the effective indices of the waveguide modes are practically equivalent at the relevant taper widths.\footnote{The only mode-dependent phase contribution is an increase of the Gouy phase shift proportional to the mode index \cite{Feng_Winful_2001}. Though inconsequential for our current beam geometries, it could be directly included as a phase correction if required in future work.}
Differences might be expected regarding optimal taper angles and the manifestation of focal shift, but in the geometries we considered these were not consequential.

In Fig.~\ref{fig:467_results_fund}b we compare the focused fundamental mode of the larger outcoupler in (a) to the HG$_{10}$ mode resulting from injecting a TE$_1$ waveguide mode into the same structure.
Interestingly, for the HG$_{10}$ emission we observed a slight improvement in outcoupling efficiency into the primary diffraction order ($\sim$1\%) due to the on-axis null, which reduces emission into the higher diffraction orders.

Adjusting the transverse focus can enable different applications.
For example, this tightly focusing outcoupler is designed to produce a strong transverse field gradient along the centerline, but other beams were designed with lobe peaks spaced 5 \textmugreek m apart for simultaneous, phase-coherent driving of neighboring ions in a string.

\subsection{Laguerre-Gaussian optical vortex beams}

Laguerre-Gaussian (LG), or ``optical vortex'', modes are another set of solutions to the paraxial Helmholtz equation, but unlike the rectangular symmetries of HG modes, these are cylindrical about the beam axis. 
The phase of a higher-order LG beam spirals about the beam axis, producing an optical vortex with topological charge $l$ which corresponds to the integer number of full phase rotations about the center in a plane normal to the beam \cite{Shen_Wang_Xie_Min_Fu_Liu_Gong_Yuan_2019}.\footnote{We restrict our consideration to the case where the radial mode index $p$ is zero, so we do not discuss it here \cite{Cui_Hui_Ma_Zhao_Han_2020}.}
With an intensity null along the beam axis and carrying $l$ quanta of orbital angular momentum, LG beams have many applications in the fields of cold atoms and trapped ions, including optical trapping and altering selection rules for electronic excitation \cite{Zhang_Robicheaux_Saffman_2011, Schmiegelow_Schmidt-Kaler_2012, Schmiegelow_Schulz_Kaufmann_Ruster_Poschinger_Schmidt-Kaler_2016,Schulz2020}.

Gratings emitting LG modes have previously been designed for IR wavelengths \cite{Nadovich_Kosciolek_Jemison_Crouse_2016, Nadovich_Jemison_Kosciolek_Crouse_2017, Zhao_Fan_2020, Zhou_Zheng_Cao_Zhao_Gao_Zhu_He_Cai_Wang_2019, Cai_Wang_Strain_JohnsonMorris_Zhu_Sorel_OBrien_Thompson_Yu_2012}.
Recently, a similar holographic approach was implemented to design gratings emitting focused free-space LG beams \cite{Zheng_Zhao_Zhang_2023}, but various approximations limit both the accuracy and generality of the design process.
Focused emission was strictly in the vertical direction, without direct control of the focused waists, and at focal heights only a few microns above the surface. 
With minor extensions detailed below, our methodology enables designs for LG beam emission without restrictions on taper geometry, beam angle, focused waists, or focal position. 

For brevity, we only present back-emitting Si$_3$N$_4$ gratings at 732 nm, but a similar variety of forward-focusing Al$_2$O$_3$ outcouplers were designed at 532 nm and produced comparable results.

\begin{figure}[!t]
\centering
\includegraphics[width=3.3in]{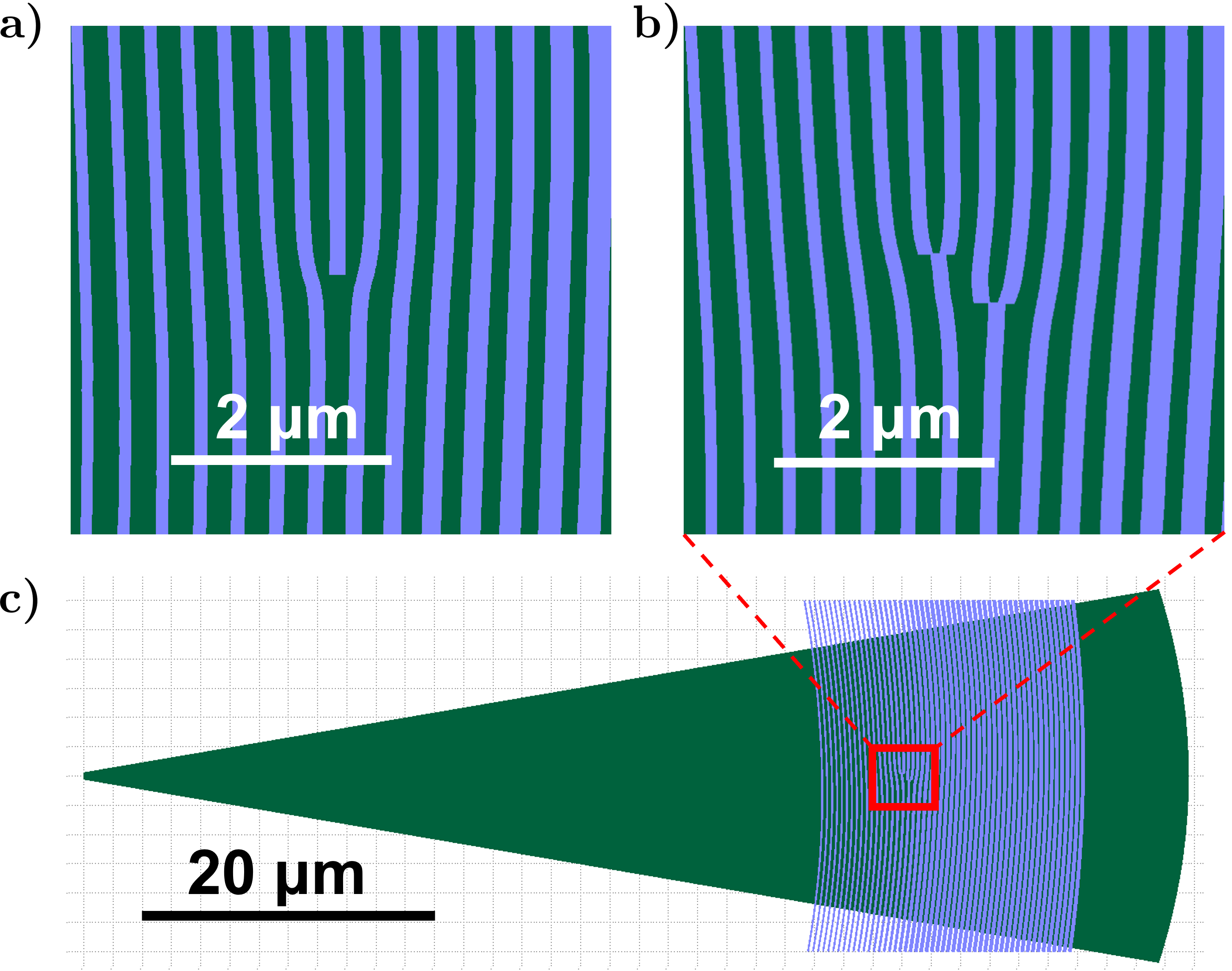}
\caption{Gratings used for non-focused emission of higher-order Laguerre-Gaussian beams (simulated beam output shown in Fig.~\ref{fig:732_LG_nonfoc_results}). 
The same taper and grating dimensions were used for both the first and second order emission, only differing in the holographically defined grating line curvatures. 
The etched lines are indicated in violet.
Zoomed regions on the beam axis center show the phase dislocations which give rise to the optical vortices. 
}
\label{fig:732_LG_nonfoc_gratings}
\end{figure}

\subsubsection{Design adaptations}

To reverse-propagate the desired beam, we extend the analytical treatment of LG beam refraction in \cite{Cui_Hui_Ma_Zhao_Han_2020} to focused beams.
The longitudinal grating parameters are still calculated from the fundamental Gaussian mode, thus the features of the higher orders are produced by the holographic phase matching---the phase contours alone will produce the central singularity and the circular intensity profile.
Our results support this approach for first and second-order LG beams ($l=1,2$), but a more involved treatment will be required at significantly higher orders to maintain sufficient mode overlap.

The phase accumulation of $2\pi l$ about the singularity forces an asymmetry in the grating plane (Fig.~\ref{fig:732_LG_nonfoc_gratings}), and as a result we no longer describe the grating lines with polynomial fits.
More importantly, however, the necessary discontinuities provide scattering points which can significantly impact the emitted beam.
In this work we make no special considerations toward handling them but discuss their impact and possible remedies below.

The current material stack (namely waveguide thickness and etch depth) was chosen to maximize grating strengths for high-efficiency emission, and we have seen that fundamental mode focusing is relatively robust to the non-Gaussian longitudinal profiles resulting from lateral feature size limitations.
In contrast, the asymmetric gratings required for LG emission make the resultant field especially sensitive to these distortions, which we find to be the primary limiting factor to emission fidelity.
In the design of future devices, lower $\alpha_\text{min}$ values should be taken into account when determining the material stack, e.g. by opting for a thinner waveguide layer or only a partial etch.
The spatial offset of substrate back-reflections can also limit mode purity, so their mitigation should also inform decisions regarding the substrate and bottom oxide thickness. 
To best represent the capabilities of the design process and future devices, we omit the silicon substrate in the designs presented below, meaning the simulation region below the waveguide layer comprises strictly oxide and produces no back-reflection.



\subsubsection{Non-focused Laguerre-Gaussian emission}

\begin{figure}[!t]
\centering
\includegraphics[width=3.3in]{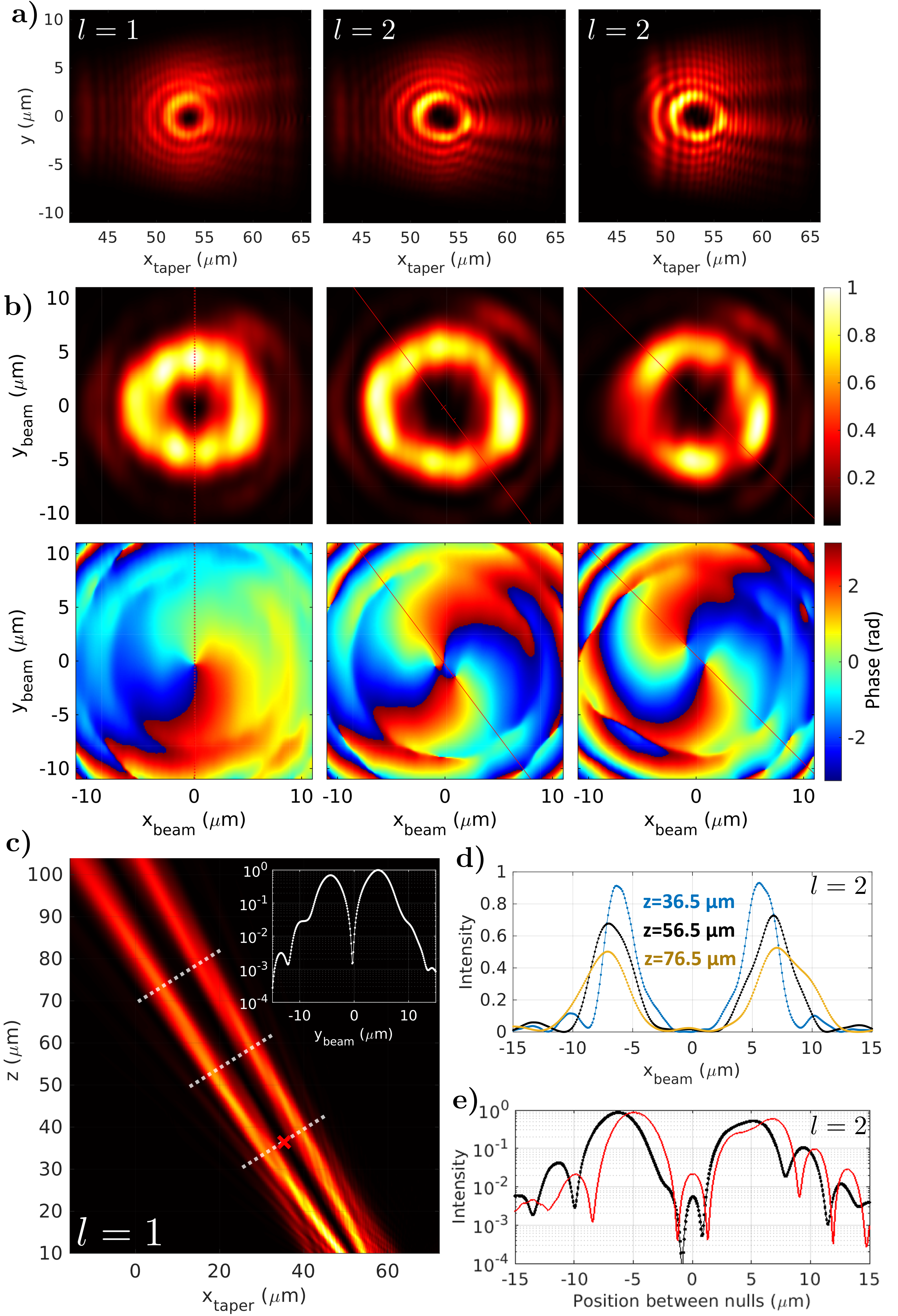}
\caption{Non-focusing Laguerre-Gaussian beam at 732 nm. Comparison of three outcouplers, all designed in Si$_3$N$_4$ for a 6 \textmugreek m waist and emission at 32\degree. From left to right in (a) and (b), the variants have minimum feature sizes of 38 nm, 38 nm, 75 nm; azimuthal orders $l=1$, $l=2$, $l=2$; and front truncation parameters 1.52, 1.52, 0.75, respectively. 
 with insets showing the intensity and phase at a beam-orthogonal slice at $z=70$ \textmugreek m (color scale same as Fig.~\ref{fig:732_LG_nonfoc_results}).
The simulated intensity is shown (a) in the plane 0.5 \textmugreek m above the oxide surface and (b) through slices orthogonal to the beam axis at 30 \textmugreek m above the chip surface ($z=36.5$ \textmugreek m), with the phase profile presented below the intensity distribution.
For comparison, intensities in (a) are normalized to the peak across all three variants; in (b) each is normalized independently. 
(c) Intensity distribution in the $xz$-plane of the $l=1$ beam. 
Inset: transverse intensity slice at $z=36.5$ \textmugreek m, along $y$ through the vortex center (the red `x' in side-view; corresponding to the vertical dotted line in (b)).
(d) Intensity cross-sections at different positions along beam path of the first $l=2$ outcoupler (minimum feature size 34 nm).
Positions of $x_\text{beam}$ slices are indicated by white dotted lines in (c).
(e) Comparison of the intensity profiles of the two $l=2$ beams along the lines connecting their split nulls in (b) (shown diagonally on plots above). 
Red corresponds to the more-truncated beam with 75 nm minimum feature size.
}
\label{fig:732_LG_nonfoc_results}
\end{figure}

Three outcouplers were designed for non-focusing LG beams at 732 nm with 6 \textmugreek m waists, emitting at 32$\degree$ from the normal.
Sharing the structure shown in Fig.~\ref{fig:732_LG_nonfoc_gratings}, these comprise one with $l=1$ and one $l=2$, each with a minimum feature size of 34 nm, and lastly an $l=2$ beam with a limit at 75 nm and appropriately chosen truncation factor.
Figure \ref{fig:732_LG_nonfoc_results} presents the resulting fields, where the smooth $l$ twists of the phase about the central axis can be clearly seen.
In the $l=2$ profiles the central vortex splits into two separate singularities.
This ``null splitting'' is a common problem in higher order LG beam generation regardless of platform \cite{Neo_Tan_Zambrana-Puyalto_Leon-Saval_Bland-Hawthorn_Molina-Terriza_2014}, and has been observed in IR coupling gratings at a comparable scale \cite{Nadovich_Jemison_Kosciolek_Crouse_2017, Zheng_Zhao_Zhang_2023}.
It is exacerbated by more stringent fabrication limitations (Fig.~\ref{fig:732_LG_nonfoc_gratings}e), as the necessary truncation brings the peak of initial emission closer to the beam center where the grating lines are increasingly asymmetric.
Methods to correct this splitting have been demonstrated and could in the future be integrated to our process \cite{Neo_Tan_Zambrana-Puyalto_Leon-Saval_Bland-Hawthorn_Molina-Terriza_2014}.

\subsubsection{Focused Laguerre-Gaussian beams}

With significantly stronger field and intensity gradients, tightly focused LG beams can provide substantial benefits in a range of applications from optical trapping \cite{Zhang_Robicheaux_Saffman_2011, Fischer_2022} to entangling gates for quantum information processing \cite{Mazzanti_Schussler_Arias, Stopp_Verde_Katz_Drechsler_Schmiegelow_SchmidtKaler_2022}.

\begin{figure}[!t]
\centering
\includegraphics[width=3.3in]{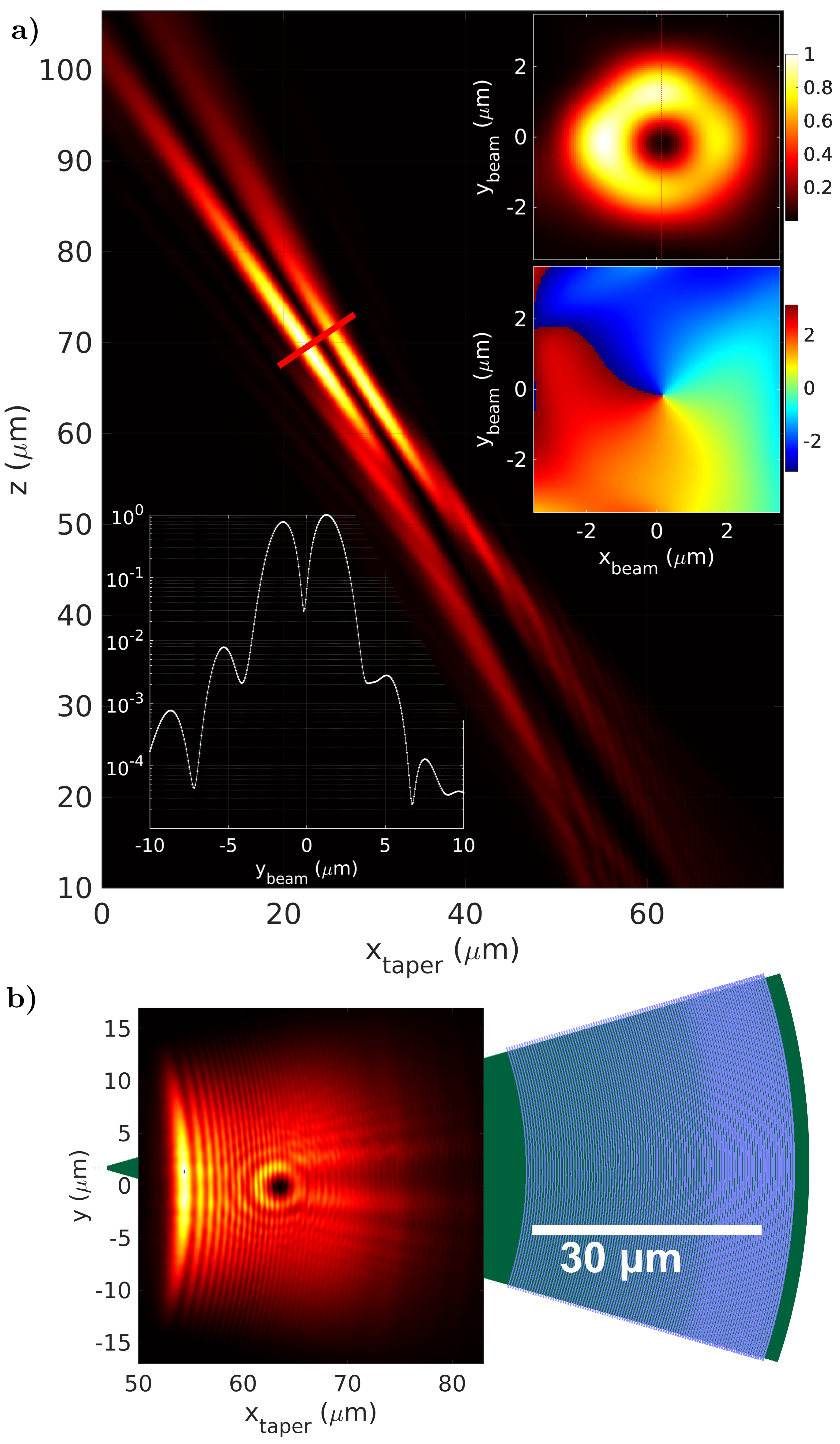}
\caption{Focused Laguerre-Gaussian emission at 732 nm. 
The beam was designed for emission at 32$\degree$, focusing to a waist of 1.5 \textmugreek m, with a minimum feature size of 75 nm, truncation at $\chi_1=0.75$ and $\chi_2=1.52$, and no silicon substrate.
(a) Intensity distribution along the $xz$-plane, with insets showing the intensity and phase at a beam-orthogonal slice at $z=70$ \textmugreek m (color scale same as Fig.~\ref{fig:732_LG_nonfoc_results}).
Inset at bottom left displays the corresponding intensity along the $y$ axis.
(b) Grating and intensity distribution at $z=7$ \textmugreek m.
}
\label{fig:732_LG_foc_results}
\end{figure}

In Fig.~\ref{fig:732_LG_foc_results} we present an outcoupler designed for focusing 732 nm light to 1.5 \textmugreek m, with designs incorporating a minimum feature size of 75 nm.
The observed focal shift agrees with that from fundamental mode focusing.
Despite the slight asymmetries in the focused profile, the outcoupling efficiency of this grating was 50\%, less than 1\% below the corresponding fundamental mode beam.

Further simulations (not shown) included the addition of a silicon substrate, which produced slight distortions to the phase profile but negligible impact on the null's intensity contrast.
More stringent fabrication limits (e.g. 150 nm) significantly perturbed the beam, leading to non-circular cross sections and intensity sometimes fluctuating by nearly an order of magnitude about the peak ring.

\subsubsection{Higher orders}
Investigations into tightly focused $l=5$ beams were marred by null splitting, the distortions appearing amplified with focusing.
Reasonable quality beams were produced for collimated emission, but the significant amount of space required for the $l$ grating line dislocations remained troublesome.
Progression to high-fidelity higher-order beams will require addressing these dislocations e.g. with phase corrections \cite{Neo_Tan_Zambrana-Puyalto_Leon-Saval_Bland-Hawthorn_Molina-Terriza_2014}.
In addition, the radius of peak intensity grows with the beam order, so it will become more critical to account for the non-Gaussian intensity profile in order to mitigate the shrinking overlap with the waveguide mode. 
Varying the period and duty cycle along the transverse direction of each grating line can enable this \cite{Zhao_Fan_2020}, but our results demonstrate that such measures are unnecessary for outcoupling into lower orders.

\section{Conclusion}

This work presents a grating outcoupler design process for waveguide-to-free-space beam-forming, capable of the precision and flexibility required for applications in trapped-ion addressing and beyond.
Ideas from previous works---namely longitudinal focusing from \cite{Mehta_Ram_2017} and holographic phase matching (e.g. as presented in \cite{Oton_2016})---were unified and built upon with a number of explicit improvements.

Firstly, rigorous analytical treatment of the desired free-space beam provides the flexibility for astigmatic, elliptical focusing while also accommodating refraction at the oxide cladding.
Secondly, explicit integration of the simulation-extracted effective index to account for the grating's influence on the propagating waveguide field enables applicability across materials and fabrication methods/grating structures.
Lastly, we utilize focal shift theory to determine the beam offset and taper angle adjustments required to ensure the desired position of the true focus.

Collectively, these improvements allow design at not only general emission angles and focal heights, but also waveguide taper geometries---such flexibility is critical in systems involving multiple such devices at different wavelengths.
Along with diffraction-limited transverse focusing in these general cases, the methodology supports the production of tightly focused Hermite-Gaussian and Laguerre-Gaussian modes.

The design process was demonstrated with a variety of outcouplers designed for trapped-ion addressing.
Device performance was analyzed with fully vectorial 3D finite-difference-time-domain simulations, including for LG mode emission.
Transverse focal shifts and their dependence on taper angle were shown to be accurately predicted by theory in both backward- and forward-emitting gratings.
The influence of minimum feature sizes was investigated, and while focused fundamental modes are relatively forgiving of distorted emission profiles, emitted LG beams were more sensitive due to the asymmetries they impose on the gratings.

Our results highlight the methodology's versatility.
The ability to rapidly and accurately produce outcouplers for many wavelengths and beam types will aid in the development of fully integrated optical systems.
With trapped ions, micron-scale focusing can enable new schemes for fast, high-fidelity readout \cite{Beck_2020}, and facilitate the use of focused Laguerre-Gaussian beams for quantum computation in long ion chains \cite{Mazzanti_Schussler_Arias}.
For neutral atoms, optical tweezers and a variety of blue-detuned point and volume traps \cite{Zhang_Robicheaux_Saffman_2011, Li_Zhang_Isenhower_Maller_Saffman_2012, Xu_He_Wang_Zhan_2010} made possible with flexibility in design can support the integration of optics in new generations of atom chips \cite{Keil_Amit_Zhou_Groswasser_Japha_Folman_2016}.
Stably delivered structured light fields may allow probing of new atom-light interactions and can support developments in on-chip spectroscopy and precision measurements \cite{Vasquez_Mordini_Verniere_Stadler_Malinowski_Zhang_Kienzler_Mehta_Home_2022, Stopp_Verde_Katz_Drechsler_Schmiegelow_SchmidtKaler_2022, Furst_Yeh_Kalincev_Kulosa_Dreissen_Lange_Benkler_Huntemann_Peik_Mehlstaubler_2020}.

\section*{Acknowledgment}

We thank LioniX International for fabrication of designs described here, and Tanja Mehlst{\"a}ubler for discussions on application of Hermite-Gauss modes to precision metrology with single ions. We acknowledge funding from ETH Zurich, the ETH/PSI Quantum Computing Hub, the Swiss National Science Foundation (Grant No. 200020\_207334), the EU Horizon 2020 FET Open project PIEDMONS (Grant No. 801285), and Cornell University. 

\ifCLASSOPTIONcaptionsoff
  \newpage
\fi



%

\bibliographystyle{ieeetr}
\bibliography{IEEEabrv,bibliog}




%


\end{document}